\documentclass[prl,twocolumn,unsortedaddress,amsmath,nofootinbib,floatfix]{revtex4-1}
\usepackage{amssymb}
\usepackage{amsmath}
\usepackage{graphicx}
\usepackage{amsbsy}
\usepackage{color}

\begin{document}
\title{Osmotic pressure induced coupling between cooperativity and stability of a helix-coil transition}

\author{Artem Badasyan}
\email{abadasyan@gmail.com}
\affiliation{Department of Theoretical Physics, J. Stefan Institute,\\ Jamova 39, SI-1000 Ljubljana, Slovenia, EU}

\author{Shushanik Tonoyan}
\affiliation{Department of Molecular Physics, Yerevan State University,\\ A.Manougian Str.1, 375025, Yerevan, Armenia}

\author{Achille Giacometti}
\affiliation{Dipartimento di Scienze Molecolari e Nanosistemi, Universit\`a Ca' Foscari Venezia,
Calle Larga S. Marta DD2137, I-30123 Venezia, Italy, EU}

\author{Rudolf Podgornik}
\affiliation{Department of Theoretical Physics, J. Stefan Institute and Department of Physics, Faculty of Mathematics and Physics, University of Ljubljana - SI-1000 Ljubljana, Slovenia, EU}

\author{V. Adrian Parsegian}
\affiliation{Department of Physics, 1126 Lederle Graduate Research Tower (LGRT), University of Massachusetts, Amherst, MA 01003-9337 USA}

\author{Yevgeni Mamasakhlisov and Vladimir Morozov}
\affiliation{Department of Molecular Physics, Yerevan State University,\\ A.Manougian Str.1, 375025, Yerevan, Armenia}

\date{\today}

\begin{abstract}

Most helix-coil transition theories can be characterized by three parameters: energetic, describing the (free) energy cost of forming a helical state in one repeating unit; entropic, accounting for the decrease of entropy due to the helical state formation; and geometric, indicating how many repeating units are affected by the formation of one helical state. Depending on their effect on the helix-coil transition, solvents or co-solutes can be classified with respect to their action on these parameters. Solvent interactions that alter the entropic cost of helix formation by their osmotic action can affect both the stability (transition temperature) and the cooperativity (transition interval) of the helix-coil transition. A consistent inclusion of osmotic pressure effects in a description of helix-coil transition, for poly(L-glutamic acid) in solution with polyethylene glycol, can offer an explanation of the experimentally observed linear dependence of transition temperature on osmotic pressure as well as the concurrent changes in the cooperativity of the transition.

\end{abstract}


\maketitle
The helix-coil transition is central to many processes in living matter \cite{cantor,molbiol}. To mimic the interactions and structures found in nature, {\sl in vitro} experiments related to biopolymers are usually performed in solutions of different composition. Naturally, a theoretical description of solvent composition effects is a necessary component of any helix-coil transition theory. There are several approaches that offer such a description within the standard Zimm-Bragg model \cite{zb,polsher}. Farago and Pincus \cite{faragopincus} proposed a classification of solvents based on their action on the parameters within this model. They have shown that solvents that promote the relative stability of the helical state compared to the coil rescale the parameter $s$, altering the melting temperature, while solvents that affect the helix-coil interfacial free energy modify $\sigma$, changing the melting interval. 
\begin{figure}[!ht]
\begin{center}
\includegraphics[width=7.3cm]{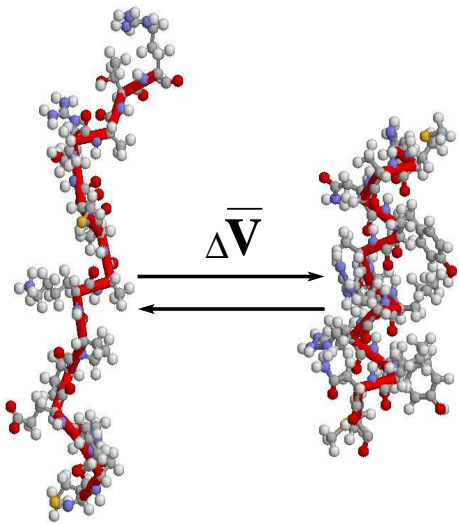}
\caption{\label{f1} Two fragments of 15 repeat units (1-15 of 2JU4 and 16-30 of 1BBA) from Protein Data Base structures are shown to illustrate the change in solvent accesible volume $\Delta\overline V$ between disordered and $\alpha$-helical conformations.} 
\end{center}
\end{figure}
Recent circular dichroism measurements of the helix-coil transition of poly(L-glutamic acid) (PLGA) in an aqueous solution of co-solvent polyethylene glycol (PEG) \cite{peg} show that besides stabilizing helices, which agrees with stability-altering solvent description of Farago and Pincus, the increase of PEG concentration also couples with the decrease of cooperativity, which is harder to explain within the Zimm-Bragg approach or its modifications \cite{faragopincus}. The main action of PEG is to modify the solvent-accessible volume within the polypeptide (see Fig.~\ref{f1}), and can thus fundamentally change the polypeptide conformational space and the associated conformational entropy. It is this coupling between the osmotic action of the PEG and the modifications on the polypeptide conformational space that will be elaborated in what follows. Here we propose a consistent theoretical framework in which a single osmotic action of  PEG solution, encoded in the osmotic pressure variation of the free energy difference between the helical and coil repeating units, engenders changes both in the melting temperature and in the melting interval, thus correctly describing the experimental data.

A minimal formulation of the Zimm-Bragg model implies a characteristic equation of the form $ (1-\lambda)(s-\lambda)-s\sigma=0 $, where the two phenomenological parameters are defined as $s=\exp(\frac{\Delta H - T\Delta S}{T})$, and $\sigma$ is the temperature-independent entropic cost of creating a helix domain within coil regions \cite{polsher}. The difference of enthalpies $\Delta H=H_{helix}-H_{coil}$ accounts for the relative energetic gain of helix formation, and $\Delta S=S_{helix}-S_{coil}$ describes the entropic cost of helix formation, with $T$ the absolute temperature, measured in units of $k_B$. The helix-coil transition occurs at the point where the energetic gain is compensated by the entropic loss, i.e., $s=\exp(\Delta H/T)/\exp(\Delta S)\sim 1$, resulting in the melting temperature $T_m \sim\Delta H/\Delta S$. The interval of transition can be estimated as $\Delta T \sim \sqrt\sigma$. Within the Zimm-Bragg formulation, the solvent effects that alter the parameter $s$ would lead to changes in the stability of the helix, modifying its transition temperature $T_m$,  while the solvent affecting $\sigma$ would alter only the cooperativity of transition $\Delta T$ \cite{faragopincus}. In this context, $s$ and $\sigma$ are introduced as completely independent parameters corresponding to two different mechanisms of solvent action. Any solvent effects that simultaneously affect both transition stability and cooperativity would then require more than a single mechanism of action, which is hard to imagine for such simple solute as PEG. Even within this Zimm-Bragg parametrization, one might suspect that the relative entropic cost of helix formation $\Delta S$ and the entropic cost of domain formation $\sigma$ could be in principle related, since they both include the entropy difference between the repeating units in the helical and coil conformations. 

The helical structure of biopolymers is mainly stabilized by intermolecular hydrogen bonding between repeating units, the presence of hydrogen bonds being a prerequisite for helix formation. A statistical description of the solvent effects on the helix-coil transition requires at least three parameters: an energetic parameter, $W=V+1=\exp(U/T)$, where $U$ is the energy of a hydrogen bond; an entropic parameter, $Q$, that stands for the ratio between the number of all accessible states versus the number of states available for the repeating unit in a helical conformation; and a geometric parameter, $\Delta$, that describes the sequential geometry of hydrogen bond formation. 
\begin{figure}[!ht]
\begin{center}
\includegraphics[width=5cm]{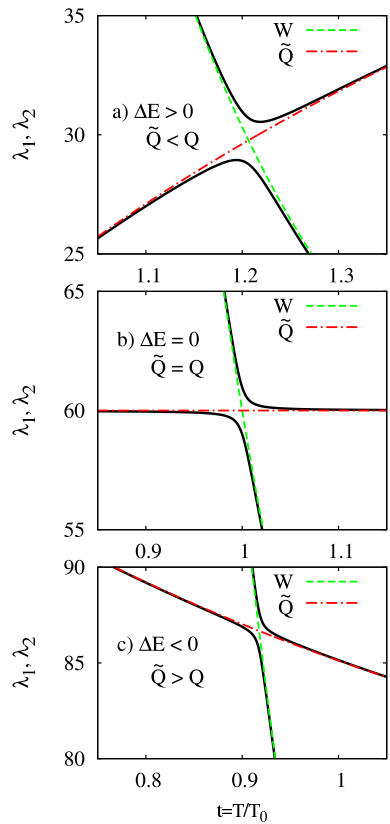}
\caption{\label{f2} Temperature dependence of two largest eigenvalues at different $\Delta E=E_h-E_c$; $Q=60$, $\Delta=3$, $q=3$. $T_0=U/\ln Q$ is the melting temperature of solvent-free model.}
\end{center}
\end{figure}
We first describe the helix-coil transition {\sl in vacuo}, i.e., without any solvent. The Hamiltonian of a solvent-free model \cite{biopoly1,biopoly2} can be written in the form
\begin{equation}
\label{ham-basic}
-\beta H_0\left(\{\gamma_i\}\right)=J\sum\limits_{i=1}^{N}\delta _{i}^{(\Delta )}
\end{equation}
\noindent as considered in \cite{biopoly1}. Here $\beta=T^{-1}$, $N$ is the number of repeating units, and $J=U/T$ is the temperature-reduced energy of hydrogen bonding, $\delta_{j}^{(\Delta )}=\prod_{k=0}^{\Delta-1}\delta (\gamma_{j+k},1)$, where $\delta (x,1)$ stands for the Kronecker symbol and $\gamma_{l}=1,\ldots,Q$. The spin variable $\gamma$ describes the state of each repeating unit by assigning to it one of the $Q$ possible conformations: conformation 1 corresponds to the helical, and the remaining $Q-1$ conformations to the coil state. The partition function 
\begin{equation}
\label{partfuncbasic}
\begin{gathered}
Z_{0}(V,Q) =\sum\limits_{\left\{ {\gamma _i=1} \right\}}^{Q} \prod\limits_{i = 1}^N  \left[ {1+V\delta _i^{\left( \Delta  \right)}} \right],
\end{gathered}
\end{equation}
can be evaluated by applying the transfer matrix method, corresponding to Eq.~\ref{ham-basic} (see Ref.~\cite{biopoly1}), resulting in the characteristic equation
\begin{equation}
\label{chareq}
\lambda^{\Delta-1}(\lambda-W)(\lambda-Q)=(W-1)(Q-1).
\end{equation}
In the thermodynamic limit, the problem simplifies considerably. It is enough to study the two largest eigenvalues obtained from Eq.~\ref{chareq}. Numerically they are closest at the point where asymptotes $W(T)$ and $Q$ cross (Fig.~\ref{f2},b), a point defining the transition temperature $T_0$. This is in accord with general physical considerations: transition occurs when entropy and energy compensate each other, at $T_0=U/\ln Q$. The minimal distance between the two eigenvalues can be estimated as $Q^{1-\Delta}$ (see \cite{biopoly2}) and is related to the helix-coil transition interval, a cooperativity measure. To quantify the transition interval, it is appropriate and informative to introduce the spatial correlation length  $\xi=\ln^{-1}\left(\frac{\lambda_1}{\lambda_2}\right)$, where $\lambda_1$ and $\lambda_2$ are the first and second leading eigenvalues of the characteristic equation. The temperature dependent $\xi(T)$ has a maximum at the transition point. Its value is related to the transition interval as $\Delta T \sim \xi_{max}^{-1} \sim Q^{\frac{1-\Delta}{2}}$ \cite{biopoly1,biopoly2}. 

Up until this point we have not taken into consideration any specific co-solute or solvent effects. Solvents able to promote solvent-polymer hydrogen bonds have been analyzed in Refs.~\onlinecite{biopoly1,bad11}. Here, however,  we consider another vast group of co-solutes which do not affect hydrogen bonding directly, but do modify the polypeptide conformations by changing the chemical potential of the solvent or the osmotic pressure of the solution. A classical example of such a co-solute is PEG, which can act as an osmoticant and as a depletion agent \cite{Depletion}. Because of their size, PEG molecules are depleted near the polypeptide chain, exerting an osmotic pressure that changes the energetic cost of certain conformations vs. others. These effects of PEG are well documented and have been explored extensively \cite{Parsegian}. The osmotic pressure of the solution depends only on the concentration of PEG, provided that all the other components can equally access the helix and the coil state of the polypeptide. It is a known function of its concentration that has been investigated in detail \cite{Cohen}. 

Our model of solvent is based on the following assumptions: i) solvent molecules can interact with (affect) both helical and coil repeating units of the polypeptide; ii) binding of solvent molecules changes the free energy of a repeating unit, depending on the conformation of the repeating unit ($E_h$ for helix, and $E_c$ for coil); iii) polypeptide-solvent interaction depends on the orientation of solvent molecules around the repeating unit and the number of solvent orientations $q>2$ account for the solvent entropy; iiii) a spin variable $\mu_i \in [1,q]$ describes the state (orientation) of a solvent molecule and its value 1 corresponds to solvent binding. 

While the various contributions to $E_h$, $E_c$ are difficult to disentangle, the overall difference $\Delta E=E_h-E_c$ can be analyzed explicitly. The (free) energy difference $\Delta E$ describes the effect of the co-solute, and the larger this difference, the stronger the stabilization of the helical state vs. the coil state. Since the main action of PEG is depletion-induced osmotic pressure, this free energy difference or equivalently the corresponding osmotic work required to drive the chain through the helix-coil transition can be written as \cite{Osmotcipress} 
\begin{equation}
\vert \Delta E\vert =  \Pi_{osm} ~\Delta \overline V.
\end{equation}
Here $\Pi_{osm}$ its the osmotic pressure of PEG solution, and $\Delta \overline V$ is the volume of water that must be exchanged with the bulk when the polypeptide chain goes through the transition \cite{Zimmerberg} (see Fig.~\ref{f1}).  This means that the helix-coil transition responds in a way analogous to a semi-permeable membrane, excluding solvent from certain portions of the polypeptide chain. The sign of $\Delta E$ depends specifically on the details of the osmotic action of PEG, stabilizing or destabilizing the helix vs. the coil state. 

We now add these solvent-mediated changes to the free energy of a repeating unit in the original {\sl in vacuo} Hamiltonian $H_0$ of the helix-coil transition \cite{biopoly1,biopoly2,bad10,bad11}. This results in
\begin{equation*}
-\beta H\left( \{\gamma_i,\mu_i\} \right)= -\beta H_0\left(\{\gamma_i\}\right) - \beta H_{solv}\left( \{\gamma_i\},\{\mu_i\} \right)
\end{equation*}
\begin{equation}
\label{hamtotal}
= \sum\limits_{i = 1}^N {\left(J\delta _i^{\left( \Delta  \right)}  + I_c\left( {1 - \delta_i^{(1)} } \right) \delta \left( {\mu _i ,1} \right)+I_h\delta_i^{(1)} \delta \left( {\mu _i ,1} \right)\right)},
\end{equation}
\noindent where $I_{c,h}=E_{c,h}/T$. The second and third terms on the rhs of Eq.~\ref{hamtotal} describe solvent interactions with repeating units in the coil and helical conformations, respectively. The  partition function is then
\begin{equation}
\label{partfunctot}
Z(V,I_h,I_c,Q) = \sum\limits_{\left\{ {\gamma _i=1} \right\}}^{Q} \sum\limits_{\left\{ {\mu _i=1} \right\}}^{q} { \exp \left( - \beta H(\{\gamma _i, \mu _i\})\right)}.
\end{equation}
\noindent Each of the solvent degrees of freedom, $\mu_i$, is coupled with a single $\gamma _i$ and the lattice expansion for Eq.~\ref{partfunctot} is of a "decorated lattice" type \cite{latt}. This allows one to sum out the solvent degrees of freedom. After a re-arrangement that makes use of the Kronecker delta properties (for details see \cite{latt}), Eq.~\ref{partfunctot} becomes
\begin{equation}
\label{partfuncrenorm}
\begin{gathered}
Z(V,I_h,I_c,Q) = (q+e^{I_h})^N Z_{0}(V,\widetilde{Q}),
\end{gathered}
\end{equation}
where
\begin{equation}
\label{Qtilde}
\widetilde{Q}=1+(Q-1)\frac{e^{I_{c}}+q-1}{e^{I_{h}}+q-1}.
\end{equation}
\noindent When Eq.~\ref{partfuncrenorm} is compared with Eq.~\ref{partfuncbasic}, it becomes obvious that the two expressions are identical up to an irrelevant multiplicative factor. Thus we can conclude that the only effect of the osmotic action of PEG is the transformation of the entropic parameter $Q \rightarrow \widetilde{Q}$. Therefore, the same characteristic equation, Eq. \ref{chareq}, but with a redefined $\widetilde{Q}$ remains valid: $\lambda^{\Delta-1}(\lambda-W)(\lambda-\widetilde{Q})=(W-1)(\widetilde{Q}-1)$. Since $\widetilde{Q}$ is now temperature dependent, the $W\sim \widetilde{Q}$ condition occurs at temperatures different from the melting temperature of the original model $T_0$. In fact, two regimes of temperature dependence of $\widetilde{Q}$ are possible, as shown in Figs.~\ref{f2}, a) and c), depending on the sign of $\Delta E$. Assuming first that the PEG osmotic pressure stabilizes the helix, $\Delta E \leq 0$, the helix-coil transition then takes place at higher temperatures (Fig.~\ref{f2}, a)). However, this is not the only effect of the PEG osmotic action: the distance between eigenvalues also becomes larger and the transition interval thus decreases. For $\Delta E \geq 0$, the situation is reversed: asymptotes cross at lower temperatures (destabilisation) and the transition interval increases (Fig.~\ref{f2}, c)). What is interesting is that even after the rescaling of parameter $Q \longrightarrow \widetilde{Q}$, Eq. \ref{Qtilde}, the point of closest approach between the two largest eigenvalues can still be estimated from the crossing between $W$ and $\widetilde{Q}$ curves. 

\begin{figure}[!ht]
\begin{center}
\includegraphics[width=7cm]{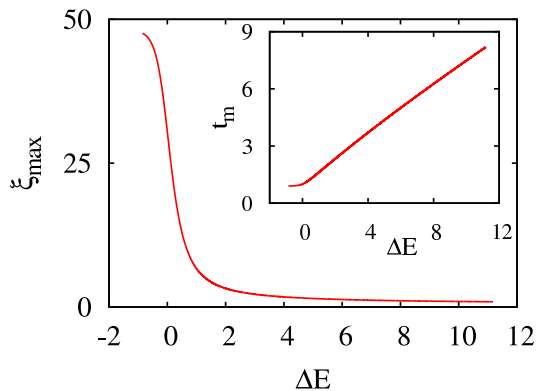}
\caption{\label{f3} The cooperatively of the transition quantified by $\xi_{max}$ as a function of $\Delta E \sim \Pi_{osm}$. Inset shows the concurrent dependence of the melting temperature $t_m = T_m/T_0$.}
\end{center}
\end{figure}

To study the stability and cooperativity of this generalized model, we analyze the temperature dependence of the correlation length from the computed eigenvalues. This allows us to calculate the transition point ($T_m$) and the cooperativity ($\xi_{max}$) as a function of  $\Delta E$ (Fig.~\ref{f3}).
As shown in Fig.~\ref{f3} (see inset), the helix-coil melting temperature grows linearly with increased $\Delta E$, indicating the increased stability of the system. This is consistent with the experimentally observed linear increase of the transition temperature with osmotic pressure as reported by Stanley and Strey \cite{chris,peg}. On the other hand, the cooperativity measure $\xi_{max}$ decreases at the same time (Fig.~\ref{f3}), as has been observed by Koutsioubas et al. \cite{peg}. This two-fold action of the PEG cannot be captured by the Zimm-Bragg model \cite{faragopincus,peg}, in which the stability and the cooperatively of the transition are described by two independent parameters.

Why can't the Zimm-Bragg based approach describe a decrease of cooperativity concurrently with an increase in stability? As  shown in \cite{biopoly2}, the Zimm-Bragg parameters can be recast in terms of our parameters as $s=W/Q$ and $\sigma=Q^{1-\Delta}\sim \xi^{-2}$. The entropic parameter $Q$ is present in both $s$ and $\sigma$. Obviously its changes  will alter both the stability and the cooperativity of the system. The assumption of independence of $s$ and $\sigma$, an inherent  property of Zimm-Bragg and related models, leads to a description of the effects of PEG \cite{faragopincus} that cannot easily be reconciled with experiments Ref.~\cite{peg}. We showed above that the resolution of this discrepancy is impossible within the Zimm-Bragg model and that an alternative, more detailed microscopic description of the PEG action, is necessary, which in its turn leads to a picture consistent with experimental results.

\begin{acknowledgments}
Authors are grateful to Valerie Parsegian for stylistic corrections she has kindly suggested. AB and AG acknowledge the support from PRIN-COFIN 2007 grant. RP and AB acknowledge ARRS grants P1-0055 and J1-4297.
\end{acknowledgments}
\bibliographystyle{apsrev}

\begin{thebibliography}{99}

\bibitem{cantor}
C. Cantor, T. Shimmel, {\it Biophysical Chemistry} (Freeman and
Co., San-Francisco, 1980).

\bibitem{molbiol}
B. Alberts, D. Bray, J. Lewis, K. Roberts and D. Watson, {\it
Molecular Biology of the Cell, Vol.1} (Garland Publ. Inc., New
York, London, 1983).

\bibitem{zb}
B. H. Zimm and J. K. Bragg \textit{J. Chem. Phys.}, \textbf{31}, 526 (1959).

\bibitem{polsher}
D. Poland, H. Scheraga, {\it The Theory of Helix-Coil Transition} (Academic Press, New York, 1970).

\bibitem{faragopincus}
O. Farago and P. Pincus \textit{Eur. Phys. J. E}, \textbf{8}, 393 (2002).

\bibitem{peg}
A. Koutsioubas, D. Lairez, S. Combet, G. C. Fadda, S. Longeville and G. Zalczer, arXiv:1112.4676v1 (2011).

\bibitem{chris}
C.B. Stanley and H.H. Strey, \textit{Biophys J}, \textbf{94}, 4427 (2008).


\bibitem{biopoly1}
N. Ananikyan, Sh. Hayryan, E. Mamasakhlisov, V. Morozov, \textit{Biopolymers} \textbf{30}, 357 {1990}.

\bibitem{biopoly2}
Sh. Hayryan, E.Mamasakhlisov, V. Morozov, \textit {Biopolymers} \textbf{35}, 75 {1995}.

\bibitem{bad10}
A. Badasyan, A. Giacometti, Y. Sh. Mamasakhlisov, V.F. Morozov and A.S. Benight \textit{Phys.Rev.E}, \textbf{81}, 021921 (2010).

\bibitem{bad11}
A. Badasyan, Sh. A. Tonoyan, Y. Sh. Mamasakhlisov, A. Giacometti, A.S. Benight and V.F. Morozov \textit{Phys Rev E}, \textbf{83}, 051903 (2011).

\bibitem{latt}
M.E. Fisher, \textit{Phys. Rev. E} \textbf{113}, 969 {1959}; D.R. Nelson and M.E. Fisher, \textit{Ann. Phys.} \textbf{91}, 226 {1975}; R. E. Goldstein, \textit{Phys. Lett. A} \textbf{104}, 285 {1984}.

\bibitem{Depletion} H. N. W. Lekkerkerker and R. Tuinier, \textit{Colloids and the Depletion Interaction}, Springer; 1st Edition (2011).

\bibitem{Osmotcipress} V. A. Parsegian, R. P. Rand, and D. C. Rau, {\sl PNAS} {\bf 97}, 3987 {2000}.

\bibitem{Zimmerberg} J. Zimmerberg and V. A. Parsegian, Nature (Lond.). {\bf 323}, 36 {1986}.

\bibitem{Parsegian} V.A. Parsegian and T. Zemb, \textit{Current Opinion in Colloid \& Interface Science}
\textbf{16}, 618 {2011}.

\bibitem{Cohen} J. A. Cohen, R. Podgornik, P. L. Hansen, and V. A. Parsegian,  \textit{J. Phys. Chem. B} \textbf{113}, 3709 {2009}.

\end{thebibliography}

\end{document}